\documentclass[12pt]{article}

\usepackage[english]{babel}
\usepackage[utf8x]{inputenc}
\usepackage{amsmath}
\usepackage{graphicx}
\usepackage{hyperref}

\title{Analysis of Yelp Reviews}
\author{Peter Hajas\\ Louis Gutierrez and Mukkai S. Krishnamoorthy\\Department of Computer Science, RPI, Troy, NY}

\begin{document}
\maketitle

\begin{abstract}
In the era of Big Data and Social Computing, the role of customer reviews and ratings can be instrumental in predicting the success and sustainability of businesses. In this paper, we show that, despite the apparent subjectivity of user ratings, there are also external, or objective factors which help to determine the outcome of a business's reviews. The current model for social business review sites, such as Yelp, allows data (reviews, ratings) to be compiled concurrently, which introduces a bias to participants (Yelp Users). Our work examines Yelp Reviews for businesses in and around college towns. We demonstrate that an Observer Effect causes data to behave cyclically: rising and falling as momentum (quantified in user ratings) shifts for businesses.

\end{abstract}

\section{Introduction}

Social networks undoubtedly play an important role in influencing which businesses consumers spend their money at. Consumers consult various sites for reviews before making decisions, such as  consumer reports, Zagat Review, and other review sites, before making a choice where to spend their money. Since the advent of Web 2.0---and various social sites, forums, etc ---getting reviews and feedback has become much easier. It follows, with this recent technology, consumers face a new problem in the form of reliability and trustworthiness of reviews. There exists the real possibility that businesses could manipulate ratings and reviews---directly and/or indirectly---causing the information that consumers rely on for good advice, to be misleading.

Another aspect of this problem comes in the form of Scholarly papers, and their ratings: either by citation counts and/or reviews. There also exists the pontial for abuse of this rating system among papers \cite{scholar}, \cite{hindex},\cite{labbe} and \cite{isi}, however, this is restricted to a small academic community. Book and movie ratings are largely still done by established institutions, and thus the impact of online ratings and reviews is minimized when compared to the previous two examples \cite{nytbook}\cite{rottentomatoes}. This is exemplified by various well-established entities, such as the Oprah Book Club and the New York Times, which have an enormous amount of influence in the area of book and film reviews. Moreover, most online purchases of books are deliberate, or premeditated, when compared to ``spur of the moment'' decisions for dining choices in a college town. There have been a number of scholarly papers written extolling the virtues of on-line publication and reviews; and there are an equal number of scholarly papers extolling the virtues of traditional publications, and condeming the biased peer reviews. The arguments for traditional publication include that peer reviews cannot be replaced by blogs, twitter or anything else - because of the slow, deliberative process that emphasizes thoughtful scholarship behind traditional publication in journals.\cite{djs}

Restaurants (especially in a college town) generally have a very short life span. Unless the restaurant is a chain, almost half of restaurant businesses may not last for more than 4 years \cite{labor}. This scenario is also typical of many small businesses \cite{msnbc}. According to BBC news \cite{bbc}, only 74 companies of the S\&P 500 companies survived for more than 40 years.  With so much uncertainty in college towns, it is often difficult to for a family business to continue past one generation. Liua et al\cite{liua} analyzes fast-food restaurant franchise data as an example for Data mining on a time series. Their work examines the impact of periodic behavior on their model.

There have been a series of articles and papers on Yelp \url{http://www.yelp.com}) Reviews. Luca\cite{luca} presents two findings: (1) That a one star increase in Yelp rating increases the revenue by 5 to 9 percent.  (2) Chain Restaurant market has decreased as a result of Yelp penetration. Blanding\cite{blanding} applauds features in Yelp in which reviewers have public profiles (as in Amazon, but not in Trip Advisor \url{http://www.tripadvisor.com}). However, according to \cite{luca}, Trip Advisor provides more extensive options for sorting and categorizing reviews. A New York times blog \cite{nytimes} cites that 2.5\% of all users in March 2008 went to Yelp (\url{http://www.yelp.com}) and the traffic has quadrupled since 2007. According to Alexa (\url{http://ww.alexa.com}) Yelp is ranked 27th in US, and ranked 132 globally as of June 19, 2014. According to Google Trends \url{http://www.google.com/trends/}, interest in Yelp peaked in 2011, and as of May 2014, has an overall interest of 59.  Yelp provides a level playing field for small restaurants, which may not be able to afford paying mass advertisement. To improve the trustworthiness of the reviews, Yelp has introduced a Review Filter system \cite{jeremy}. They employ an algorithmic filter mechanism, whose purpose is to protect consumers and business owners from fake, shill and malicious reviews.
 
 In this paper we analyze the data that was supplied for twenty campus restaurant locations over a period of 7 years. We provide a simple spring (device) like model to explain the behavior of Yelp ratings from 2005 to 2011. We provide an experimental validation by taking five different restaurant locations around geographically diverse college campuses. Lastly, we explain how our model fist with the ratings that were given in Yelp reviews.



\section{Model and Data}

As previously stated, the Yelp Academic Dataset provides users reviews for 7 years from 2005 to 2011 for a select set of university campuses and neighborhood restaurants. The universities include:
\begin{enumerate}
\item Brown University
\item California Institute of Technology
\item California Polytechnic State University
\item Carnegie Mellon University
\item Columbia University
\item Cornell University
\item Georgia Institute of Technology
\item Harvard University
\item Harvey Mudd College
\item Massachusetts Institute of Technology
\item Princeton University
\item Purdue University
\item Rensselaer Polytechnic Institute
\item Rice University
\item Stanford University
\item University of California - Los Angeles
\item University of California - San Diego
\item University of California at Berkeley
\item University of Illinois - Urbana-Champaign
\item University of Maryland - College Park
\item University of Massachusetts - Amherst
\item University of Michigan - Ann Arbor
\item University of North Carolina - Chapel Hill
\item University of Pennsylvania
\item University of Southern California
\item University of Texas - Austin
\item University of Washington
\item University of Waterloo
\item University of Wisconsin - Madison
\item Virginia Tech
\end{enumerate}

In this paper first we examine how ratings influence the quality of restaurants, and the cyclical nature of this behavior. We assume that to produce quality food, one has to continue to invest money on production, service and advertisement. However once the service is established and good ratings are obtained, continued investment on the components that contribute to the quality may drop. This results in a lower rating resulting in investment, and thereby improving the quality.  

We use a differential equation to model the ratings/quality of restaurants over a period of time. The solution to the linear second order differential equation describes the various features that we have observed in reviews over the years. The collective reviews converge on a stable value (number), and individual restaurant reviews tend to demonstrate periodic behavior.  The right hand side is a demand (or the number of users) to the restaurant, and we use three constants to model damping, average number of customers and the resilience of the restaurant.

A linear second order differential equation $m \frac{d^2}{dt^2} x +c \frac{d}{dt}x + k x= q \cos(\omega t)$ with appropriate initial conditions describes the motion of a spring. $x$ is the unit stretched from normal length, $k$ is the spring constant, $m$ is the mass, $c$ the damping factor and $q \cos( \omega t)$ is a forcing function.  In this research we use a similar model to explore the relationship between the services or the food quality, denoted by  $q$ ,of a restaurant (We assume a relationship between $q$ and the ratings ). The pull or, increase in the number of the customers, puts a strain on the restaurant, just as it would a physical spring. In springs, one uses Newton's second law of motion to arrive at the second order differential equation. In restaurants, one can use the law of conservation of the number of customers $q$ in the above function. The variable in restaurants, $x$ represents the change in ratings (result of change in service/quality).  The solution of this second order differential equation is either convergent, oscillatory or divergent depending upon the values of $k$ and $m$.

In the case of restaurants, we know the solution, i.e., the rating, can never diverge (as the maximum rating is 5 and the minimum rating is 1). We predict that the individual restaurant reviews will be periodic: that the damping factor $c$ will be very small.

For example, the solution to the differential equation, for values of $m, c,k$ to be 1 and $q$ to be 10, and $\omega$ to be 5, demonstrates periodic behavior. Since each individual restaurant has shows periodic behavior, we expect that the cumulative reviews will converge. We provide only a qualitative evidence and not a quantitative evidence (by finding appropriate values for the parameters of the differential equation and initial conditions). 

However, there is an additional factor that complicates. Even though each restaurant service/rating may be modeled as a second order differential equations, these equations are coupled. This is the result of a free flow of capital, i.e., chefs, among these restaurants. In our framework, we assume that these differential equations are uncoupled.

If there are larger number of reviews/ratings for a few restaurants, we expect that the number of ratings follow a power law. This is similar to the power law feature in many social sciences networks  \cite{newman},\cite{powerlaw}.
We plot the number of reviews/ratings in log-log scale to illustrate the powerlaw.

We provide heat maps of where most reviewed restaurants are located. The aim of this is to show that most reviewed restaurants are usually clustered together. This is an intuitive results, given that consumers have the option of choosing another (equally good) restaurant if the originally chosen restaurant is crowded.

We have specifically focused on the wide fluctuation of ratings (based on quality and service) of restaurants around Troy, NY over the period of time studied.  The restaurant names, and the number of reviews for each restaurant are listed next in table \ref{tab:counts1} (sorted by the most reviewed ones first). For the sake of brevity, we have restricted the listing to only 30 names. However, our analysis includes the all the restaurants and services (a total of 158 restaurants and services). Even though we have tried to clean the data, more work is needed in this regard(i.e. some restaurant names are repeated). 
\begin{table}
 \begin{tabular}{ ||l | c ||}
\hline
Name of the Restaurant & Total Number of Reviews\\
     \hline

Browns Brewing Co.: The Taproom &76\\
Dinosaur Bar-B-Que &62\\
Famous Lunch &38\\
DeFazio's Pizzeria & 37\\
Jose Malones & 34\\
Ali Baba &29\\
The Ruck &28 \\
I Love NY Pizza& 26\\
Snowman &23\\
Shalimar Restaurant & 23\\
Flavour Cafe  Lounge& 23\\
X's To O's Vegan Bakery& 22\\
Plum Blossom Chinese Restaurant& 21\\
Ale House &21\\
Beirut Restaurant &19\\
Troy Market &18\\
Illium Cafe& 17\\
Spill'n the Beans Coffeehouse  Bistro &16\\
Daisy Baker's &15\\
Bacchus &15\\
Red Front Restrnt  Tavern &14\\
Sushi King &13\\
Country View Diner& 13\\
The Greek House &12\\
Muza &12\\
Manory's Restaurant& 12\\
Francesca's &12\\
The Placid Baker &11\\
South End Tavern &11\\
Lo Porto's & 11\\

\hline

\end{tabular}
\caption{\label{tab:counts1} Restaurants Vs. Number of Reviews (in decreasing order).}
\end{table}

\begin{figure}
\centering
\includegraphics[width=0.5\textwidth]{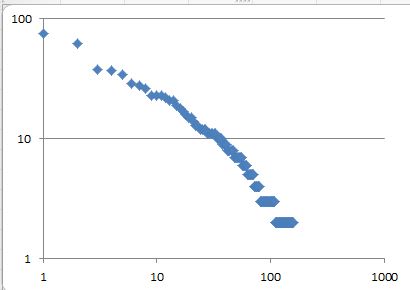}
\caption{\label{fig:fig1}{Number of ratings Vs Restaurants and Services in log-log scale}}
\end{figure}

The number of ratings appears to follow a power law curve by Figure \ref{fig:fig1} Plot.  (The X-axis represents the number of restaurants/services and the Y-axis represents the number of reviews)

Table \ref{tab:year_average} gives the average rating for all restaurants near Troy. The overall average rating converges to an average of 3.75.

\begin{figure}
\centering
\includegraphics[width=0.5\textwidth]{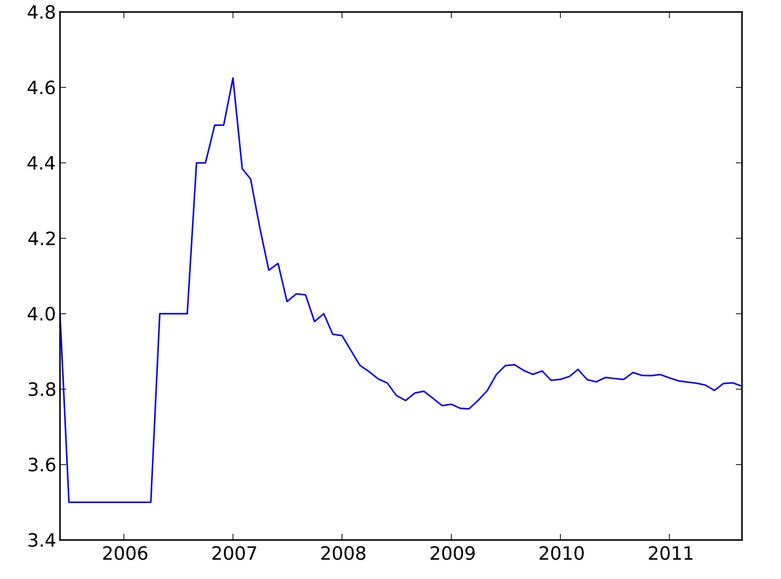}
\caption{\label{fig:fig2}{Running Average of ratings of all Restaurants and Services Vs Time Period}}
\end{figure}

 \begin{tabular}{ ||l | c ||}
\hline
Year &Average\\
     \hline
2005& 3.5\\
2006 &5.0 \\
2007 &3.85245901639\\
2008& 3.66666666667\\
2009 &3.91139240506\\
2010& 3.83435582822\\
2011&3.7660944206\\
\hline
\label{tab:year_average}
\end{tabular}
\\
\\
\noindent
Figure \ref{fig:fig2} shows the Average rating over the time period 2005 to 2011.

\begin{figure}
\centering
\includegraphics[width=0.5\textwidth]{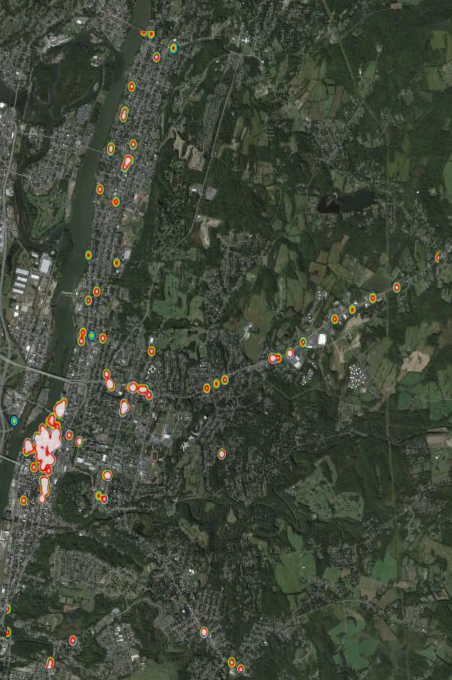}
\caption{\label{fig:fig3}{Heat Map of Restaurant in a map - based on the average Reviews}}
\end{figure}

Figure \ref{fig:fig3} shows a heat map of where all these restaurants are located.

\begin{figure}
\centering
\includegraphics[width=0.5\textwidth]{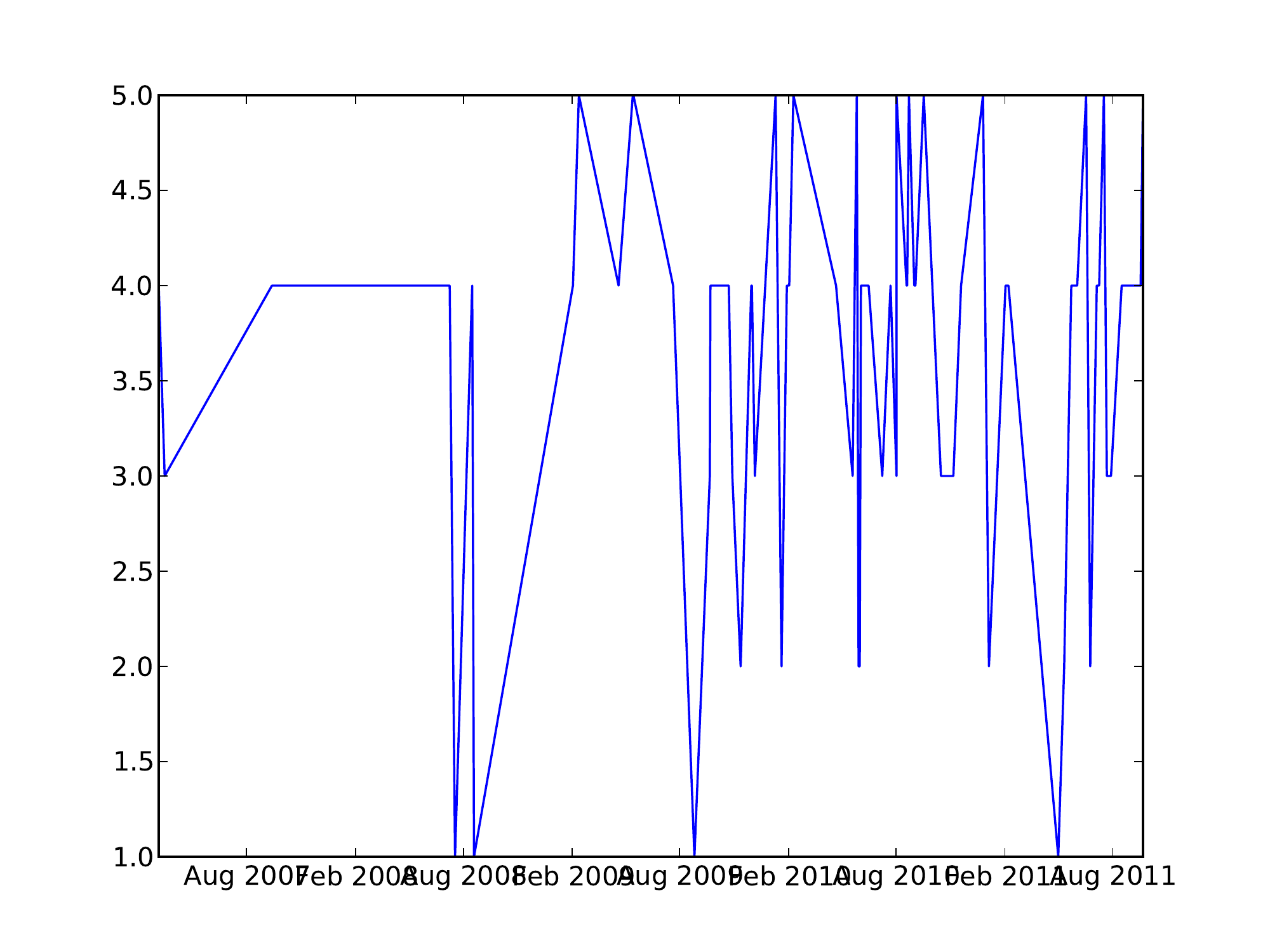}
\caption{\label{fig:fig4}{Restaurant Review of Brown Brewing Tap Room Restaurant over the time period }}
\end{figure}

Figure \ref{fig:fig4} shows the periodic behavior of reviews for the business with the maximum number of reviews, Browns Brewing Co.: The Taproom.

\begin{newpage}

\section{Analysis of Data}

The same model was applied on Yelp data from four (geographically) different college campuses. The four places analyzed are as follows:

\begin{enumerate}
\item Ann Arbor, Michigan 
\item Manhattan (near Columbia University), New York City
\item Seattle, Washington
\item Raleigh, Chapel Hill and Durham, North Carolina
\end{enumerate}

\subsection{Ann Arbor}
First we look at Ann Arbor (Home of University of Michigan) Listed below are the top restaurants and their review counts in table \ref{tab:counts2}.

\begin{table}
 \begin{tabular}{ ||l | c ||}
\hline
Name of the Restaurant & Total Number of Reviews\\
     \hline
Zingerman's Delicatessen &407\\
Ashley's &147\\
Madras Masala& 90\\
Sava's& 87\\
Eve& 87\\
Comet Coffee& 84\\
Silvio's Organic Pizza &83\\
Angelo's Restaurant& 77\\
New York Pizza Depot &76\\
Tomukun Noodle Bar &71\\
Sadako Japanese Restaurant& 66\\
Tio's Mexican Cafe &65\\
Bubble Island &62\\
Lab& 61\\
Totoro &58\\
Red Hawk Bar  Grill &58\\
Northside Grill &58\\
No Thai &58\\
Gandy Dancer Restaurant& 58\\
Good Time Charley's &51\\
Douglas J Aveda Institute Salon &50\\
Sushi.Come &49\\
Yamato& 48\\
TK Wu &48\\
Bar Louie &48\\
Yogo Bliss& 45\\
The Original Cottage Inn &45\\
Raja Rani Fine Indian Cuisine &45\\
Pita Kabob Grill &45\\
Amer's Delicatessen& 45\\
China Gate Restaurant& 44\\
Panchero's Mexican Grill& 41\\
Le Dog &41\\
Brown Jug &40\\

\hline \hline
\end{tabular}
\caption{\label{tab:counts2} Restaurants Vs. Number of Reviews (in decreasing order).}
\end{table}

The number of reviews again follow the Power Law, and is shown in figure \ref{fig:fig5}, where the X-axis represents the number of restaurants/services and the y-axis the number of reviews.
\begin{figure}
\centering
\includegraphics[width=0.5\textwidth]{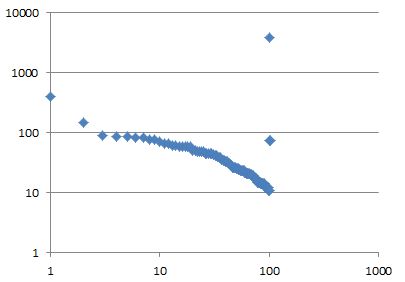}
\caption{\label{fig:fig5}{Number of ratings Vs Restaurants and Services in a log-log scale}}
\end{figure}
\noindent

Figure \ref{fig:fig6} shows the Average rating from 2005 to 2011.

\begin{figure}
\centering
\includegraphics[width=0.5\textwidth]{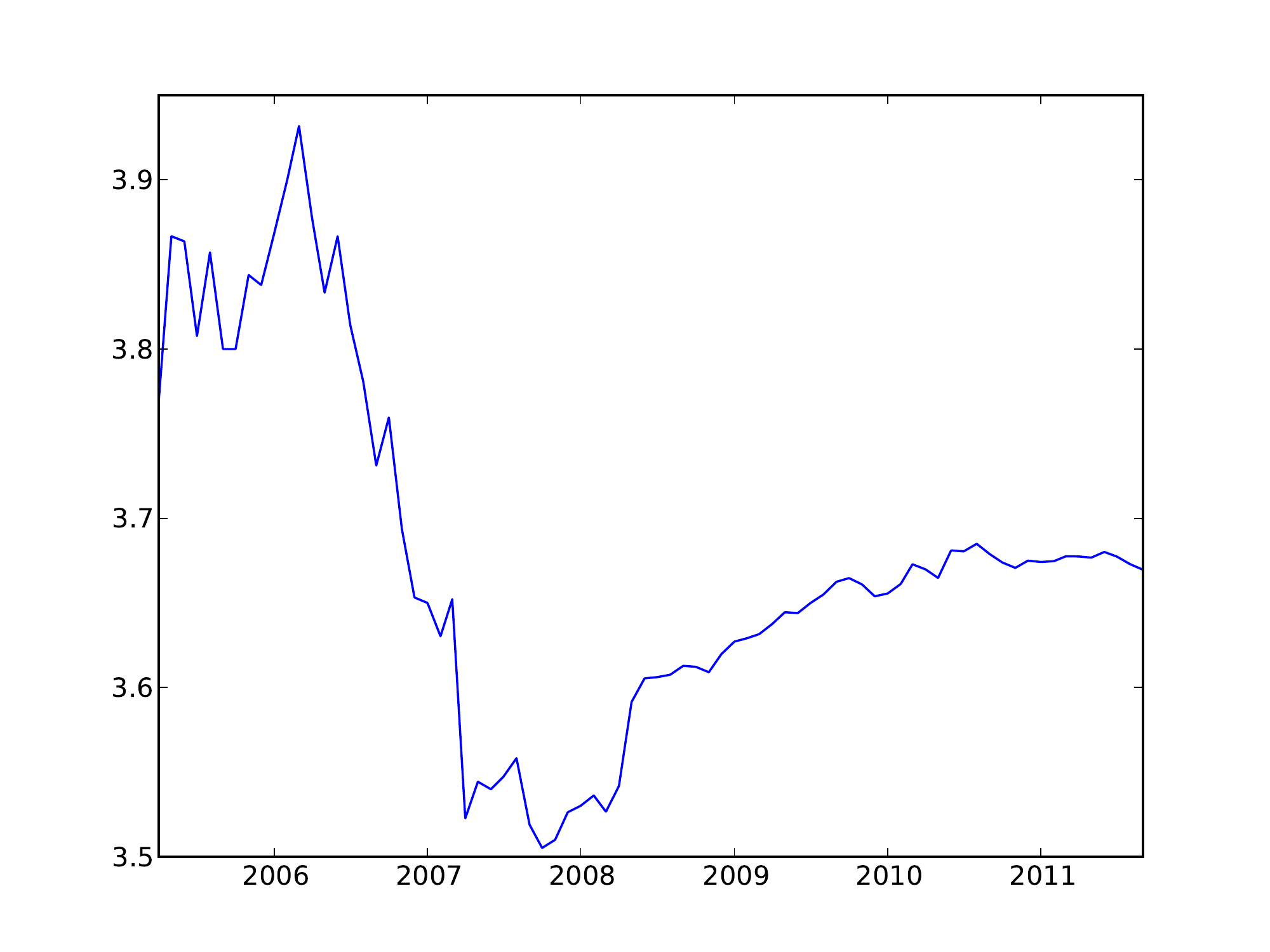}
\caption{\label{fig:fig6}{Running Average of ratings of all Restaurants and Services Vs Time Period}}
\end{figure}

The results show that the average rating converges close to 3.7. Figure \ref{fig:fig7} heat map of where all these restaurants are located.

\begin{figure}
\centering
\includegraphics[width=0.5\textwidth]{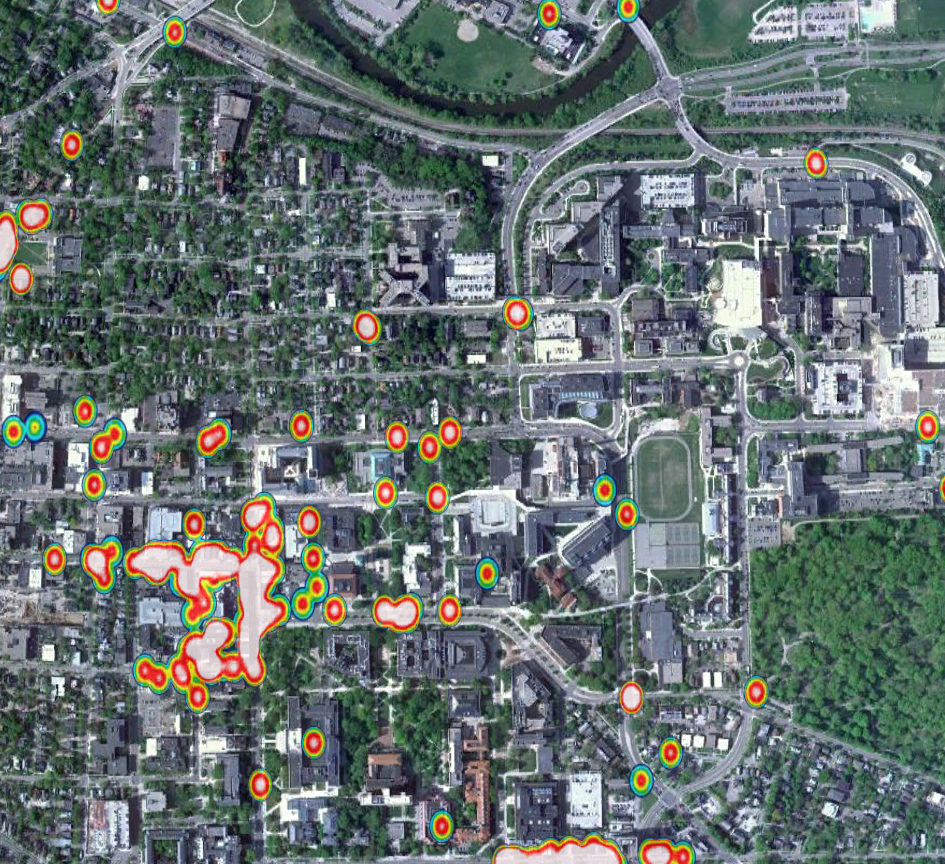}
\caption{\label{fig:fig7}{Heat Map of Restaurant in a map - based on the average Reviews}}
\end{figure}

To show the cyclic behavior of the reviews for one restaurant, figure \ref{fig:fig8} shows the most reviewed restaurant, Zingerman's Delicatessen.

\begin{figure}
\centering
\includegraphics[width=0.5\textwidth]{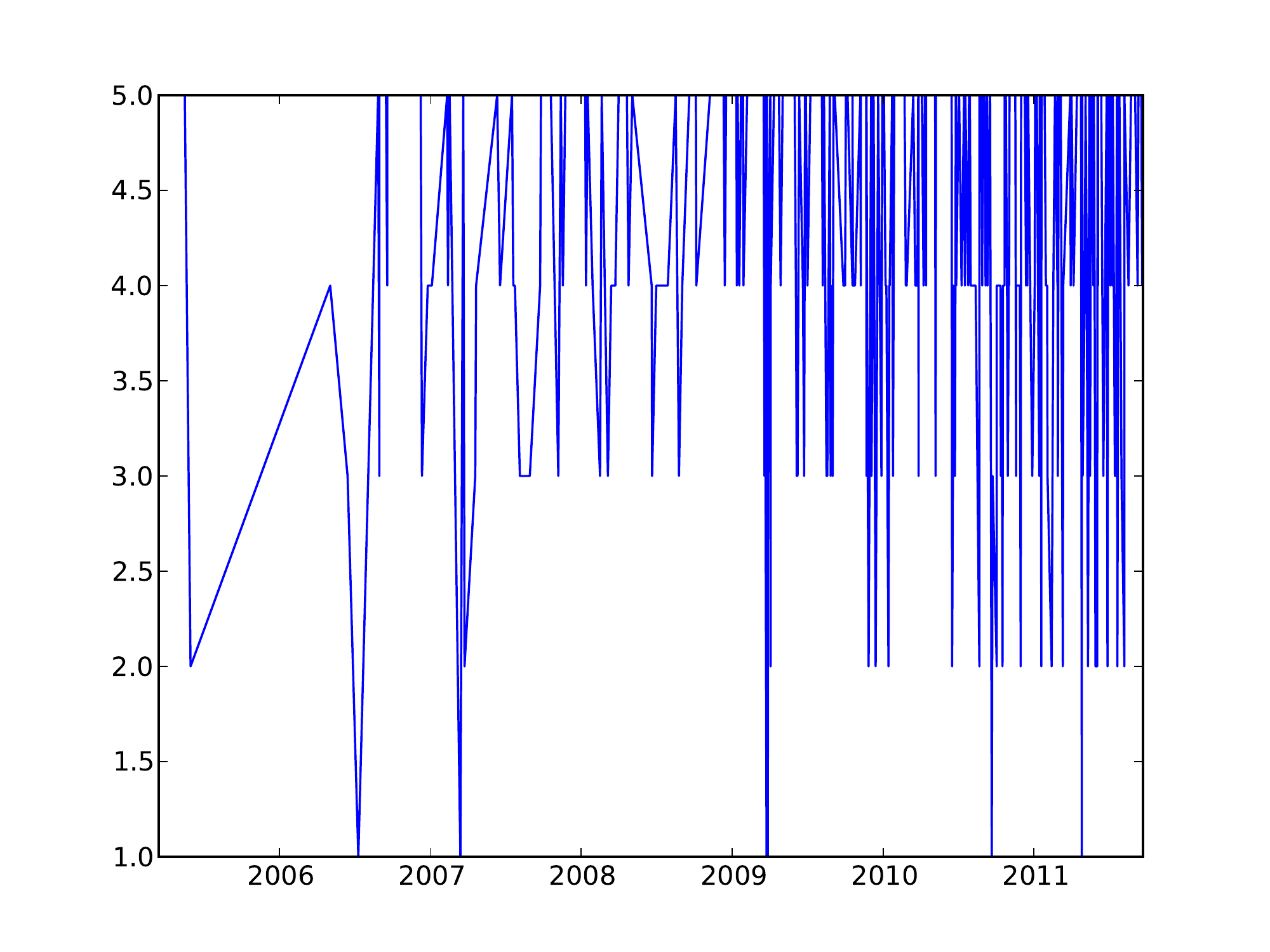}
\caption{\label{fig:fig8}{Restaurant Review of Zingerman's Delicatessen  Restaurant over the time period }}
\end{figure}

Fluctuations are obvious in the above figure, however some of the fluctuation may be due to noise. In order to reduce noise, the highest reviews were used.

\subsection{Manhattan}
Next, we look at Manhattan (Home of University of Columbia University) Listed below are the top restaurants and their review counts.
\begin{table}
 \begin{tabular}{ ||l | c ||}
\hline
Name of the Restuant & Total Number of Reviews\\
     \hline
Koronet Pizza &267\\
Community Food Juice & 265\\
Tom's Restaurant& 216\\
Kitchenette Uptown &190\\
Hungarian Pastry Shop &185\\
Max Soha &160\\
Havana Central& 155\\
Melba's& 151\\
Society Coffee& 150\\
Le Monde& 138\\
Ollie's &135\\
Miss Mamie's Spoonbread Too& 129\\
Deluxe on Broadway& 129\\
Roti Roll - Bombay Frankie& 126\\
Mel's Burger Bar& 123\\
Rack  Soul &117\\
The Heights Bar  Grill& 114\\
Sip &105\\
Vareli& 94\\
Mill Korean& 94\\
Nussbaum  Wu Bakery &88\\
M2M &88\\
Campo &87\\
Artopolis &86\\
The Village Pourhouse& 80\\
Westside Market NYC &77\\
V  T Pizzeria  Restaurant& 77\\
Harlem Tavern& 76\\
Vine: Sushi  Sake& 75\\
Lion's Head Tavern &75\\
Max Caf &73\\

\hline \hline
\end{tabular}
\caption{\label{tab:counts3} Restaurants Vs. Number of Reviews (in decreasing order).}
\end{table}

As seen on earlier two cities (Troy and Ann Arbor), the number of reviews follow a power law as shown in figure \ref{fig:fig9}, where the X-axis represents the number of restaurants/services and the y-axis represents the number of reviews.

\begin{figure}
\centering
\includegraphics[width=0.5\textwidth]{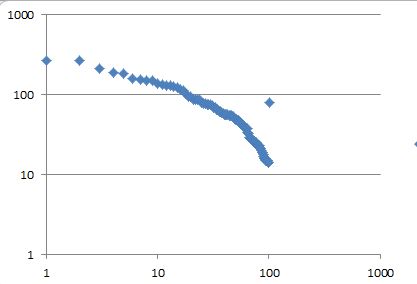}
\caption{\label{fig:fig9}{Number of ratings Vs Restaurants and Services in a log-log scale}}
\end{figure}
\noindent
Figure \ref{fig:fig10} shows the average rating over from 2005 to 2011.

\begin{figure}
\centering
\includegraphics[width=0.5\textwidth]{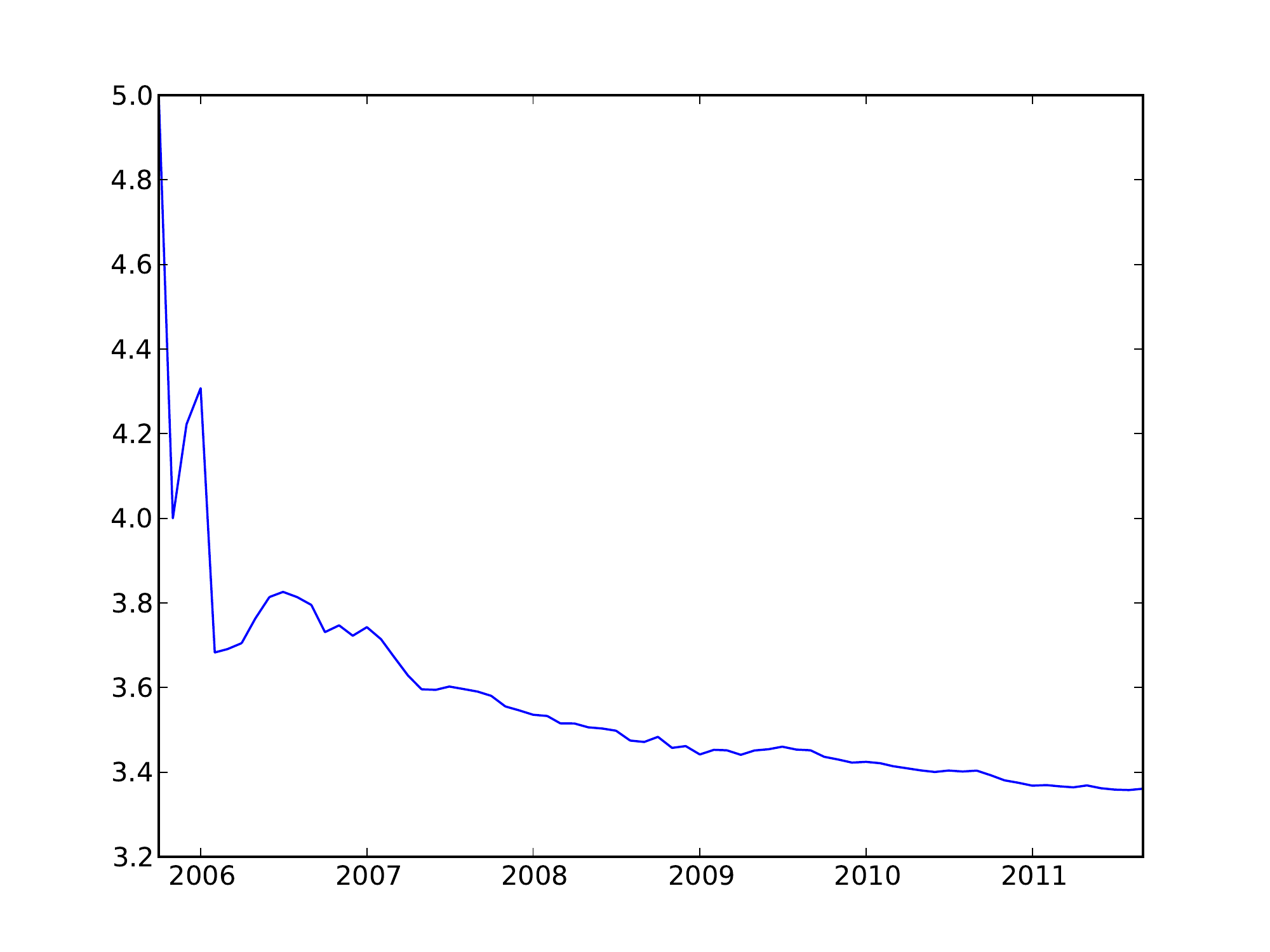}
\caption{\label{fig:fig10}{Running Average of ratings of all Restaurants and Services Vs Time Period}}
\end{figure}

The results show that the average rating converges close to 3.4 for the restaurants in Manhattan (close to Columbia University). Figure \ref{fig:fig11} shows a heat map of where all these restaurants are located.

\begin{figure}
\centering
\includegraphics[width=0.5\textwidth]{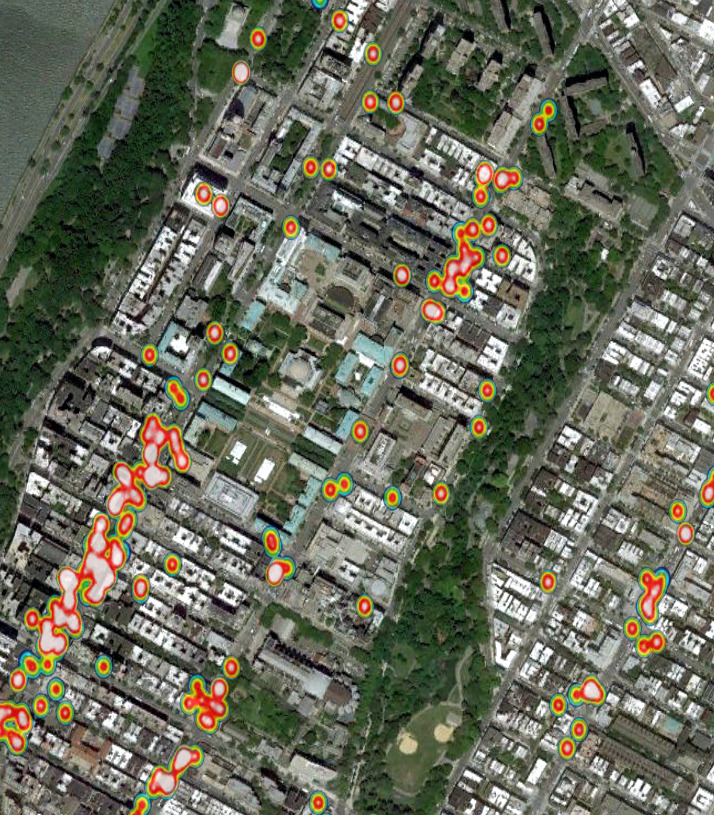}
\caption{\label{fig:fig11}{Heat Map of Restaurant in a map - based on the average Reviews}}
\end{figure}

To show the cyclic behavior of the review of one restaurant, figure \ref{fig:fig12} shows the most reviewed restaurant, Koronet Pizza.

\begin{figure}
\centering
\includegraphics[width=0.5\textwidth]{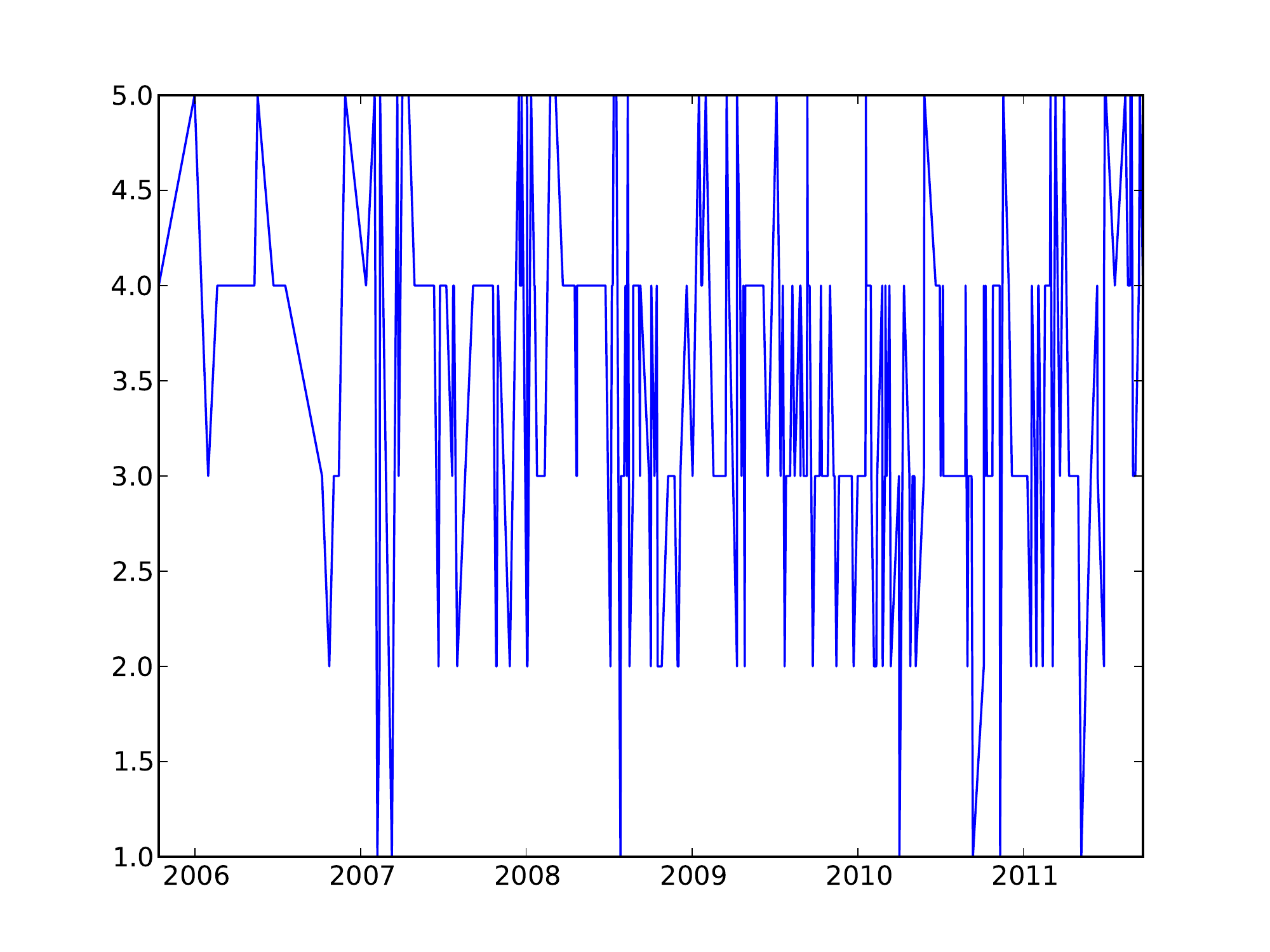}
\caption{\label{fig:fig12}{Resturanat Review of Koronet Pizza Restuarant over the time period }}
\end{figure}

\subsection{Seattle}

Next, we look at Seattle (Home of University of Washington), located in the Pacific Northwest of the United States. Listed below are the top restaurants and their review counts.
\begin{table}
 \begin{tabular}{ ||l | c ||}
\hline
Name of the Restuant & Total Number of Reviews\\
     \hline
Thai Tom &680\\
Blue C Sushi& 285\\
Aladdin Gyro-Cery &189\\
Shultzy's Sausage &166\\
Trabant Coffee  Chai& 160\\
Thanh Vi &154\\
Cafe Solstice& 148\\
Aladdin Falafel Corner& 140\\
Thaiger Room &136\\
Big Time Brewing Company &129\\
Flowers Bar  Restaurant& 128\\
Pho Than Brothers& 127\\
Ugly Mug Caf &126\\
Chipotle &125\\
Cafe On The Ave& 119\\
Samurai Noodle &115\\
Boom Noodle &111\\
Thai 65& 110\\
University Teriyaki& 108\\
Sonrisa &105\\
Yunnie Bubble Tea &102\\
WOW Bubble Tea &100\\
Continental Restaurant  Pastry Shop& 96\\
Cedars &93\\
Buffalo Exchange& 92\\
Sureshot &90\\
Jimmy John's &90\\
Fran's Chocolates &90\\

\hline \hline
\end{tabular}
\caption{\label{tab:counts4} Restaurants Vs. Number of Reviews (in decreasing order).}
\end{table}

As seen on earlier three cities (Troy, Ann Arbor and Manhattan), the number of reviews here follow a power law as shown in figure \ref{fig:fig13}, where X-axis represents the number of restaurants/services and the y-axis represents the number of reviews.

\begin{figure}
\centering
\includegraphics[width=0.5\textwidth]{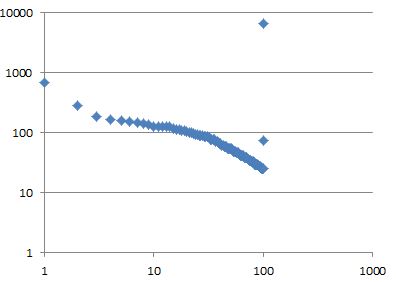}
\caption{\label{fig:fig13}{Number of ratings Vs Restaurants and Services in a log-log scale}}
\end{figure}
\noindent
Figure \ref{fig:fig14} shows the Average rating from 2005 to 2011.

\begin{figure}
\centering
\includegraphics[width=0.5\textwidth]{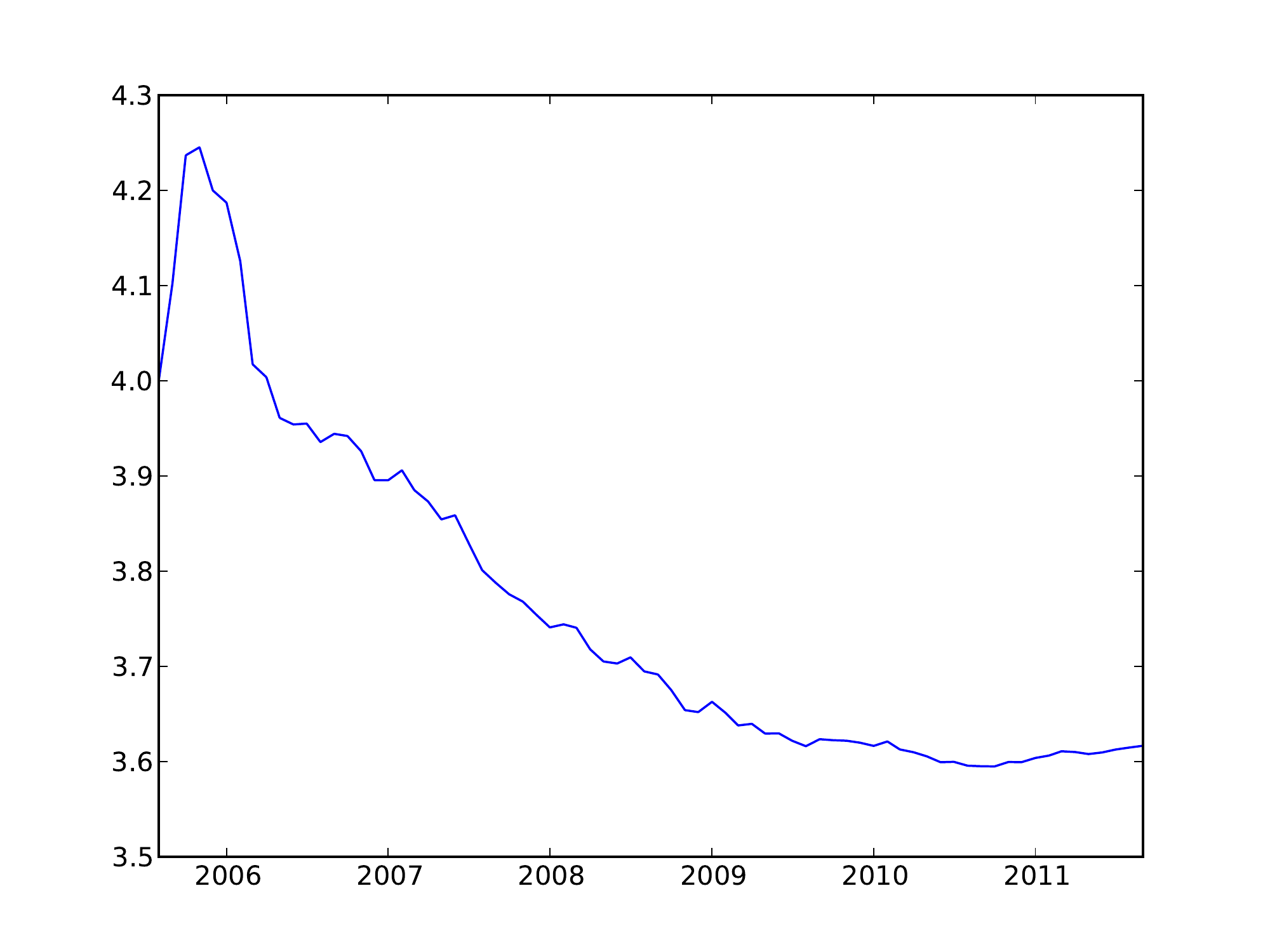}
\caption{\label{fig:fig14}{Running Average of ratings of all Restaurants and Services Vs Time Period}}
\end{figure}

The results find that the average rating converges close to 3.6 for the restaurants in Seattle area (close to University of Washington). Figure \ref{fig:fig15} shows a heat map of where all these restaurants are located.

\begin{figure}
\centering
\includegraphics[width=0.5\textwidth]{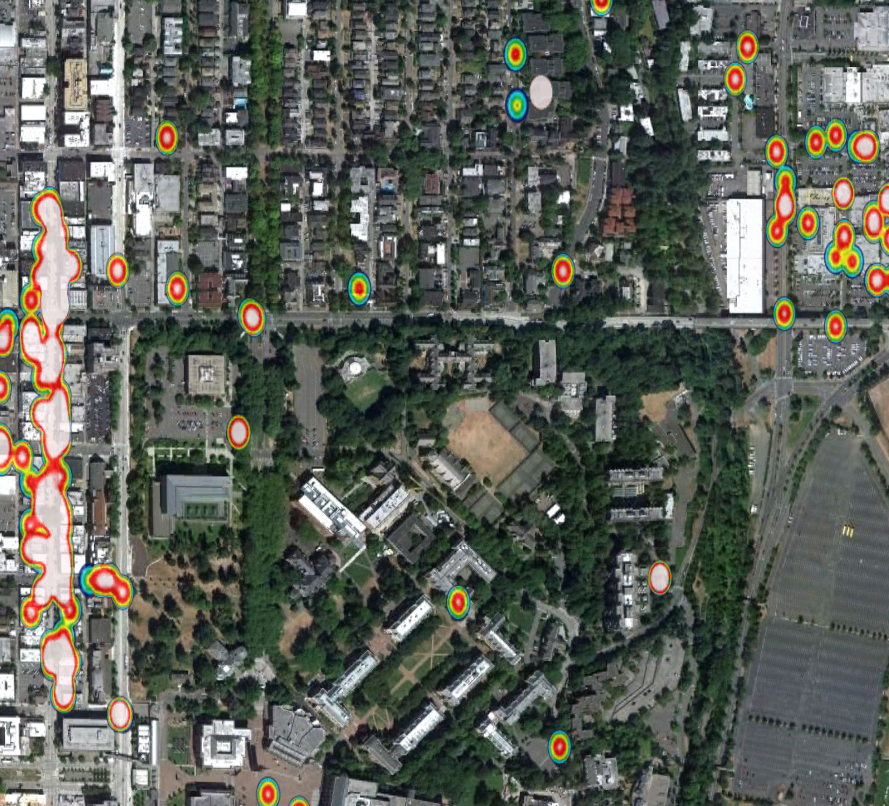}
\caption{\label{fig:fig15}{Heat Map of Restaurant in a map - based on the average Reviews}}
\end{figure}

To show the cyclic behavior of the review of one restaurant, figure \ref{fig:fig16} shows the most reviewed restaurant, Thai Tom.

\begin{figure}
\centering
\includegraphics[width=0.5\textwidth]{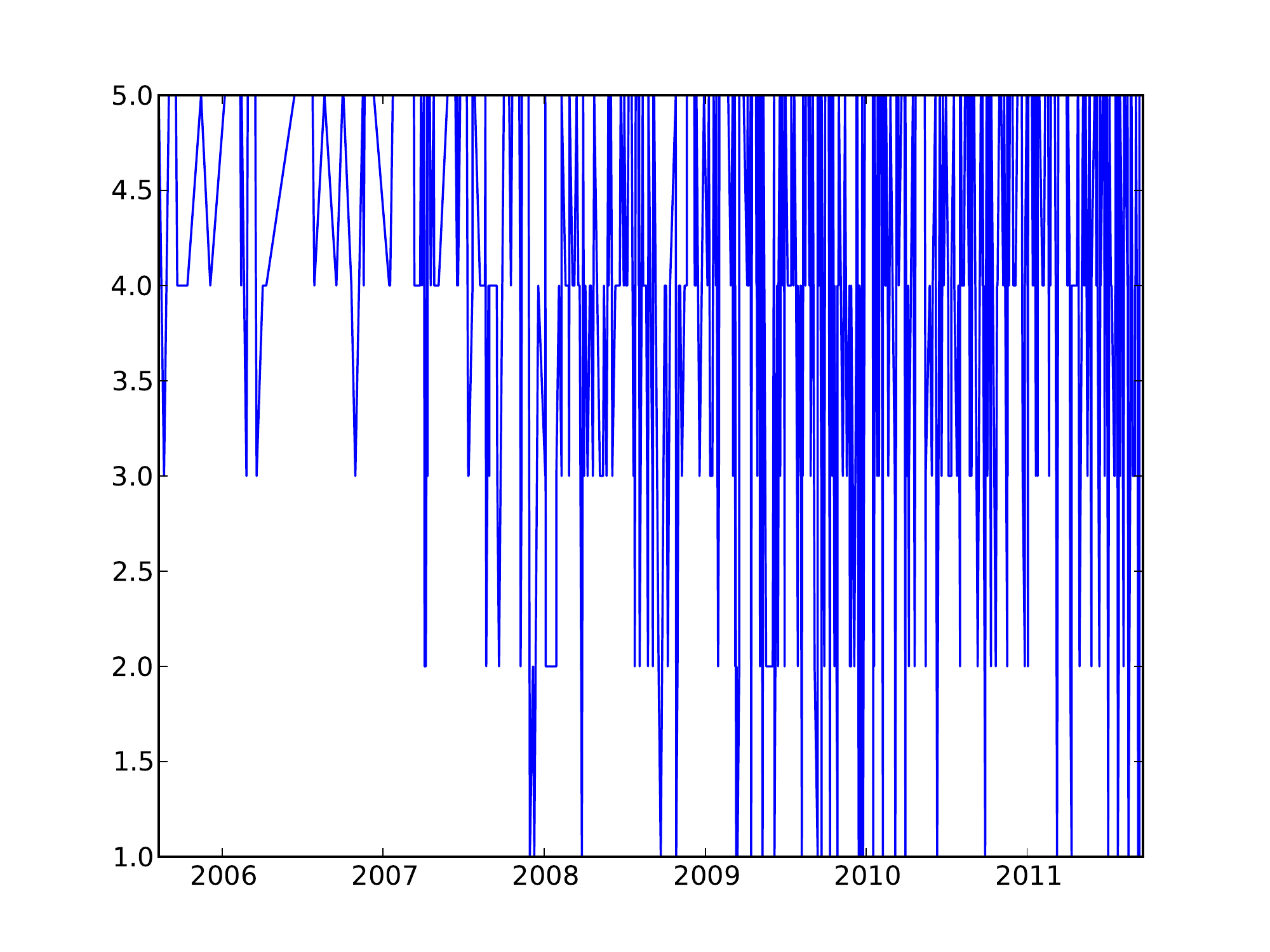}
\caption{\label{fig:fig16}{Restaurant Review of Thai Tom Restaurant over the time period }}
\end{figure}

\subsection{Raleigh, Chapel Hill and Durham}

Lastly, we look at Raleigh, Chapel Hill and Durham (Home of North Carolina State University, University of North Carolina and Duke), located in East Coast of the United States. Listed below are the top restaurants and their review counts.
\begin{table}
 \begin{tabular}{ ||l | c ||}
\hline
Name of the Restuant & Total Number of Reviews\\
     \hline
Mediterranean Deli  Catering& 122\\
Top of the Hill Restaurant  Brewery& 110\\
Lantern &95\\
Buns &92\\
Mama Dip's Kitchen& 80\\
Sandwhich &78\\
Pepper's Pizza &76\\
411 West &74\\
Mint& 70\\
Crook's Corner& 63\\
Lime and Basil& 61\\
Carolina Brewery& 59\\
Sugarland& 57\\
Vimala's Curryblossom Cafe& 54\\
Time-Out Restaurant& 52\\
Elaine's On Franklin& 48\\
The Crunkleton &47\\
Ye Olde Waffle Shoppe &45\\
Penang& 43\\
Cosmic Cantina &43\\
Foster's Market &40\\
35 Chinese Restaurant& 40\\
Talullas &39\\
Italian Pizzeria III& 34\\
Yogurt Pump &33\\
Sutton's Drug Store& 33\\
Spanky's Restaurant &32\\
Local 506 &32\\
Linda's Bar and Grill& 32\\
Moshi Moshi &31\\
Jack Sprat Cafe &30\\

\hline \hline
\end{tabular}
\caption{\label{tab:counts5} Restaurants Vs. Number of Reviews (in decreasing order).}
\end{table}

The table below gives the average rating for all restaurants near Raleigh, Chapel Hill and Durham Area. The overall average rating converges to 3.75.

 \begin{tabular}{ ||l | c ||}
\hline
Year &Average\\
     \hline
2005 &4.22222222222\\
2006 &4.13157894737\\
2007 &3.75630252101\\
2008 &3.93542757417\\
2009 &3.76438356164\\
2010 &3.80676328502\\
2011 &3.7443324937\\

\hline
\end{tabular}

As seen on earlier four cities (Troy, Ann Arbor, Manhattan and Seattle), the number of reviews for this area follow the power law as shown in figure\ref{fig:fig17}, where X-axis represents the number of restaurants/services and the y-axis represents the number of reviews.
\begin{figure}
\centering
\includegraphics[width=0.5\textwidth]{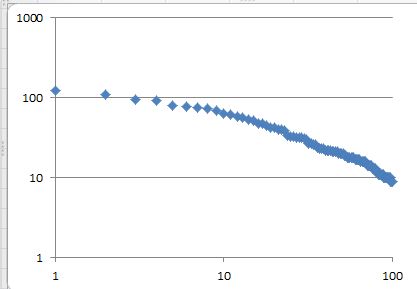}
\caption{\label{fig:fig17}{Number of ratings Vs Restaurants and Services in a log-log scale}}
\end{figure}
\noindent
Figure \ref{fig:fig18} shows the Average rating from 2005 to 2011.

\begin{figure}
\centering
\includegraphics[width=0.5\textwidth]{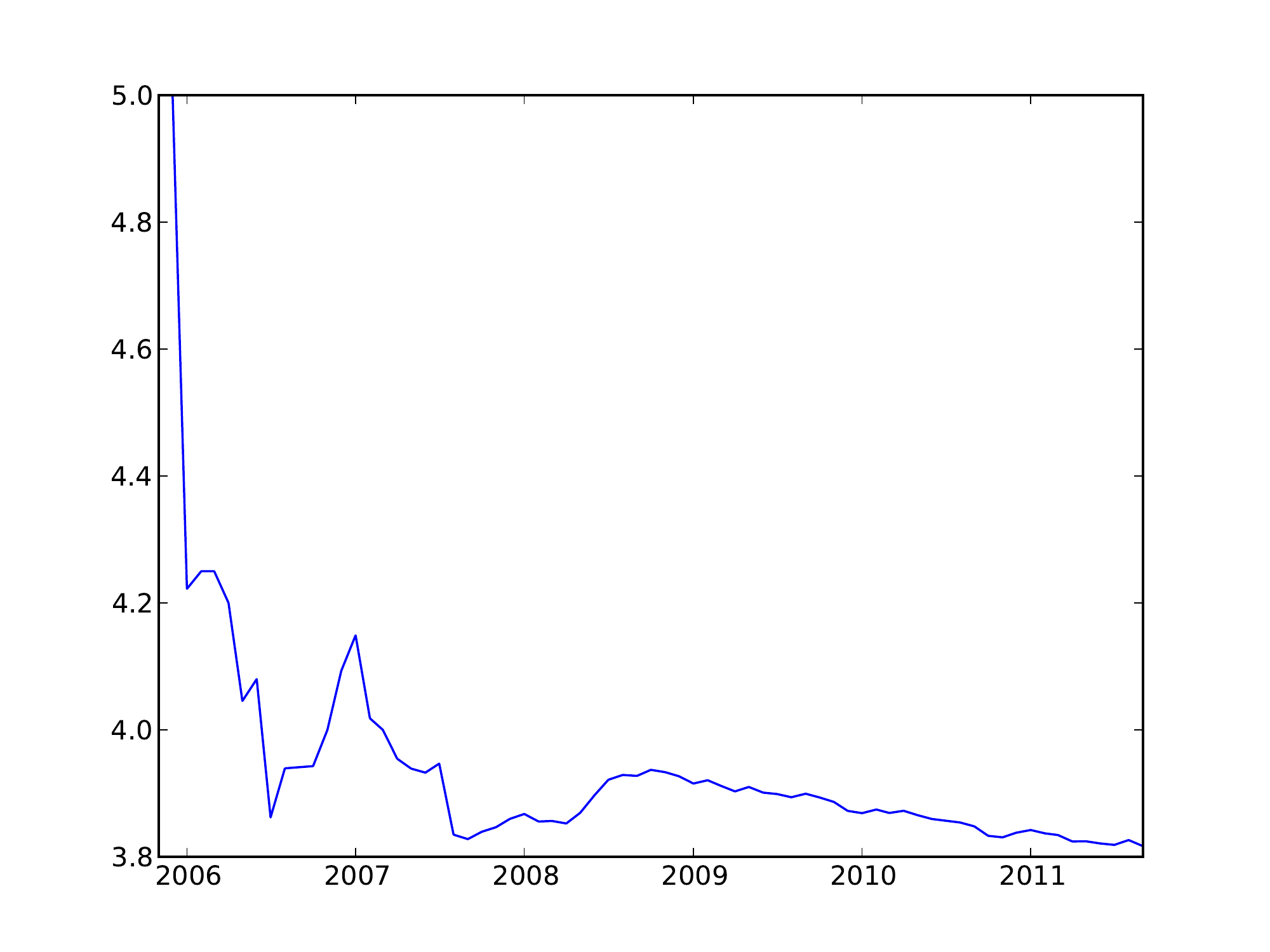}
\caption{\label{fig:fig18}{Running Average of ratings of all Restuarants and Services Vs Time Period}}
\end{figure}

The results show the average rating converges close to 3.8 for the restaurants in Raleigh, Chapel Hill and Durham area (close to three Universities). Figure \ref{fig:fig19} shows a heat map of where all these restaurants are located.

\begin{figure}
\centering
\includegraphics[width=0.5\textwidth]{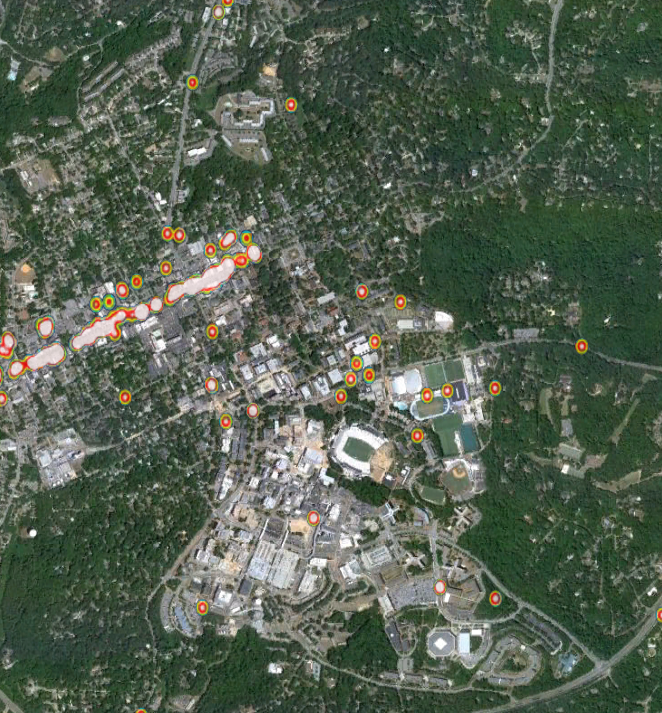}
\caption{\label{fig:fig19}{Heat Map of Restaurant in a map - based on the average Reviews}}
\end{figure}

To show the cyclical behavior of the reviews of one restaurant, figure \ref{fig:fig20} shows the most reviewed restaurant, Mediterranean Deli and Catering.

\begin{figure}
\centering
First we look at Ann Arbor (Home of University of Michigan) Listed below are the top restaurants and their review counts.
\includegraphics[width=0.5\textwidth]{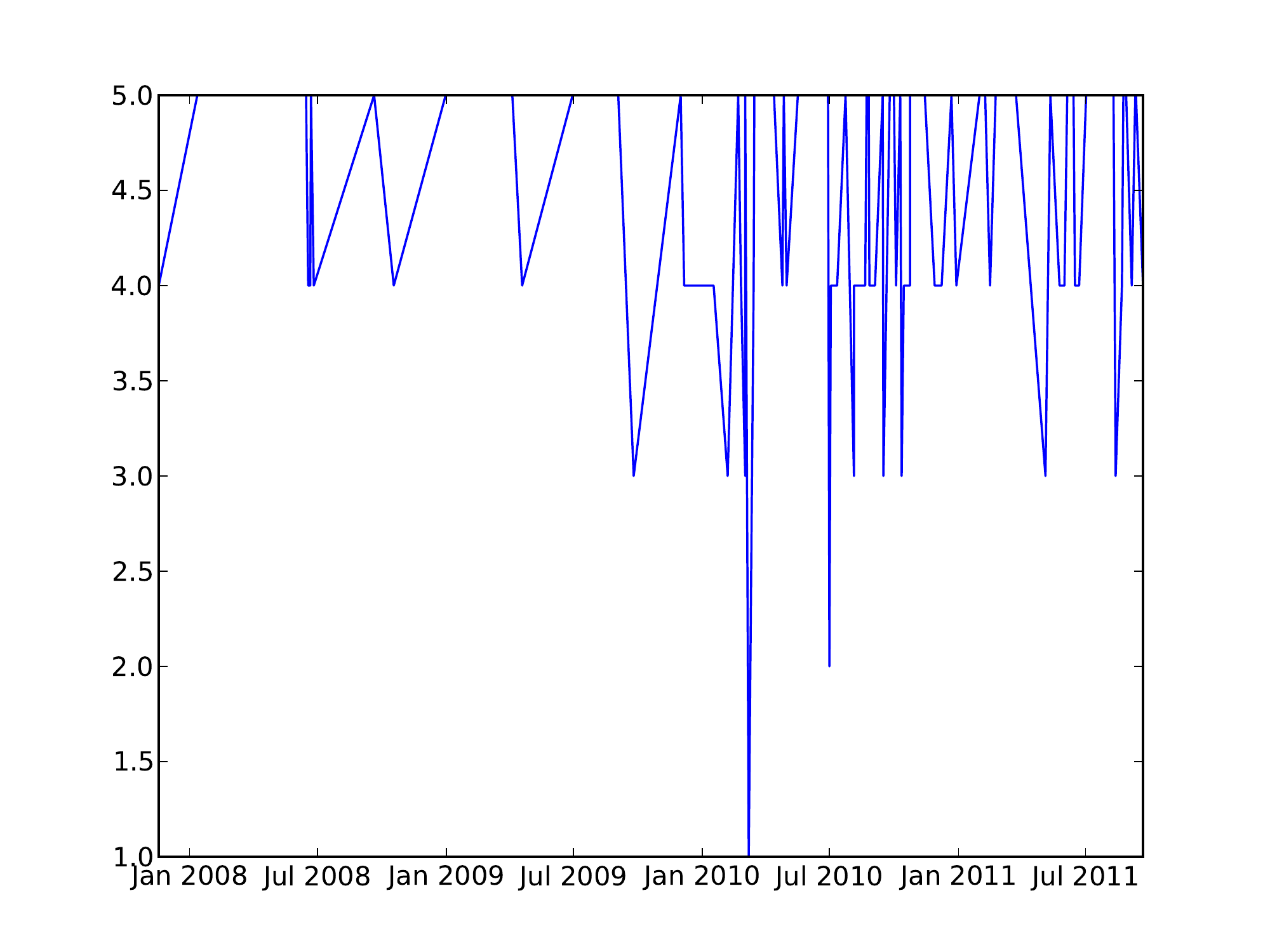}
\caption{\label{fig:fig20}{Resturanat Review of Mediterranean Deli  Catering Restuarant over the time period }}
\end{figure}

\section{Discussions}

The results presented in this paper demonstrate that the cumulative rating of restaurants as a whole converge, whereas the reviews of individual restaurants have fluctuating behavior. This phenomenon may also be explained as historical patterns repeat themselves. This phenomenon may also be explained by food preferences of college students, such as the preference for ethnic restaurants, and how top ethnicity food preferences may differ depending upon the regions. Pizza restaurants also appear as some of the most frequently rated restaurants, and by and large, they are the most common among all the campuses. Even though university sponsored restaurants appear in all the lists, they do not tend to be frequently reviewed. Frequently reviewed restaurants appear to be clustered close to each other. This may be due to attracting the overflow crowd, or just that these restaurants are situated in a popular area for dining.

The periodic behavior of an individual restaurant could may also be attributed to a lack of data (reviews) during a specific time period. We have posted the ratings of all (restricted to top 100) the restaurants (for the five locations we considered) in the following URL 
\url{http://goo.gl/16bcnF } as compressed zip files. 

\section{Acknowledgement}
The third author wishes to thank Prof. G. M. Prabhu and Dr. J. Krishnamoorthy for their help in editing the paper. The third author also wishes to thank Prof. H. S. Mani for his help on the model.  All three authors wish to thank Mr. Sean O'Sullivan '85 for his generous donation to the Rensselaer Center for Open Source Software where this work originated.

\end{newpage}

\end{document}